\documentclass[review]{elsarticle}
\usepackage{graphicx} 
\usepackage{amsfonts} 
\usepackage{amsmath} 
\usepackage{amsthm}	
\usepackage{textcomp} 

\journal{Reliability Engineering \& Systems Safety}
\begin{document}
\begin{frontmatter}

\title{A Graphical method for simplifying Bayesian Games}

\author[leeds]{Peter A. Thwaites\corref{cor1}}
\ead{P.A.Thwaites@leeds.ac.uk}
\author[wwk]{Jim Q. Smith}
\ead{J.Q.Smith@warwick.ac.uk}

\address[leeds]{School of Mathematics, University of Leeds, LS2 9JT, United Kingdom}
\address[wwk]{Department of Statistics, University of Warwick, Coventry, CV4 7AL, and Alan Turing Institute, United Kingdom}

\cortext[cor1]{Corresponding author}

\begin{abstract}
If the influence diagram (ID) depicting a Bayesian game is common knowledge to its players then additional
assumptions may allow the players to make use of its embodied irrelevance statements. They can then use these to discover
a simpler game which still embodies both their optimal decision policies. However the impact of this result has been
rather limited because many common Bayesian games do not exhibit sufficient symmetry to be fully and efficiently
represented by an ID. The tree-based chain event graph (CEG) has been developed specifically for such asymmetric problems. By
using these graphs rational players can make analogous deductions, assuming
the topology of the CEG as common knowledge. In this paper we describe these powerful new techniques and illustrate them 
through an example modelling a game played between a government department and the provider of a website designed to radicalise vulnerable people.

\end{abstract}

\end{frontmatter}

\section{Introduction}

There are two principal conceptual difficulties in applying results 
from Bayesian game theory in a number of domains. Firstly, whilst it
might be plausible for a player to know the broad structure of an opponent's
utility function when that opponent is subjective expected utility maximizing (SEUM), for a player to
also believe that she knows the \emph{exact} quantitative form of that
utility function or the precise formulation of the distribution of its
attributes is less plausible. Secondly, as for example \cite{Nau} has pointed out,
however compelling our beliefs are that an opponent's rationality \emph{should} induce her to be SEUM, in practice most 
people simply are not. So any application of a theory which starts with this assumption is hazardous. These issues induced \cite{KadLar} to
suggest giving up on the rationality hypothesis entirely and instead modelling
the opponent simply in terms of her past behaviour.

However others have perservered with rationality modelling by addressing these
real modelling challenges more qualitatively. For example \cite{Plausible}
suggested a way to address the first difficulty described above. We can continue to model successfully provided that 
the conditional independences associated with various hypotheses and the
attributes of each player's utility function are common knowledge, but we do not need
that the players know the quantitative forms of others' inputs. This
framework developed from methods for simplifying influence diagrams (IDs) \cite{HowardandM,JandPboth},
described first in \cite{Shachter86} and then \cite{Jim1989}. When players are
all SEUM, substantive conclusions can sometimes be made concerning those
aspects of the problem upon which a rational opponent's decision rules
might depend. This in turn allows players to determine ever simpler
forms for their own optimal decision rules. So models can be built
which at least respect some of the structural implications of rationality
hypotheses {\bf before} being embellished with further structure gleaned from
behavioural data, or the bold assumption that an opponent's quantitative
preferences and beliefs can be fully quantified by everyone. Even the second
criticism of a Bayesian approach outlined above is at least partially
addressed, since the methods need only certain {\bf structural} implications of
SEUM to be valid, not that all players are SEUM. 

In this way game theory can therefore be used not to fully specify the quantitative form of a
competitive domain but simply to provide hypotheses about the likely \emph{dependence} structure that rationality 
assumptions might imply for such models. These models can then be embellished with further historical
quantitative information using the conventional Bayesian paradigm. 

It has been possible to demonstrate the efficacy of the approach when
modelling certain rather domain-specific applications, but it has proved rather
limited in scope~\cite{Plausible,Allard}. One problem is that the structure of many games cannot
be fully and effectively represented by an ID (see for example~\cite{BandShen,CandM,QZP}). Usually the underlying game
tree is highly asymmetric and so the symmetries necessary for an
encompassing and parsimonious ID representation of the game are not
present. 
This is one characteristic of the types of games that we consider in this paper; we discuss other important attributes in the following paragraphs.

Harsanyi~\cite{Harsanyi} considered games where the players are uncertain about some or all of the following -- the other players' utility
functions, the strategies available to the other players, and the information other players have about the game. In the games considered in
this paper each player holds a body of common knowledge -- the exact form of other players' utility functions is unknown, but 
the variables these functions depend upon (a feature of the conditional independence structure of the game) are known; the strategies available
to the other players are known; and what information is known to the other players is also known. 
So essentially the games we consider are ones where the ``structure" is common knowledge, but the exact values of other players' utilities
and the probability distributions of some chance variables are not.

In contrast to Harsanyi, Banks et al in~\cite{Banks} state that
{\it Game theory needs the defenders to know the attackers' utilities and probabilities, and the attackers to know the defenders' 
utilities and probabilities, and for both to know that these are common knowledge.}
We do not agree that these are absolute requirements, but it is certainly true that a player cannot {\it solve} a game to her satisfaction unless 
she has some values for her opponents' utilities and probabilities. So in our games, players assign subjective probabilities to their {\it unknowns}
and estimate values of their opponents' utility functions. Each player's utilities depend not only on the strategies chosen by the various
players, but also on chance.

We take a decision-theoretic approach to Bayesian Game theory. Our games are sequential (typically with players acting alternately, and with chance
variables interspersed between the players' actions). The standard description for such a game is {\it Extensive Form Bayesian Game with Chance
moves}; they are generally expressed as a {\it game tree} (or as an ID~\cite{Plausible} or MAID -- multi-agent influence diagram~\cite{KolandM}).

Asymmetric games, as described above, are being played with increasing frequency wherever large constitutional organisations (governments, police forces
etc.) are at risk from or attempting to combat criminal or anticonstitutional organisations or networks. An example is described below, taken 
from this area, probably less familiar than games in a commercial context.

Governments and police play a game with groups trying to influence or radicalise susceptible individuals.
These {\it radicalisers} often
attempt to influence vulnerable people via the web. The government strategy here can be thought of as a combination of {\it prevention} and
{\it pursuit}: if a website is easily accessible then it might be best just to shut it down; if it is difficult to access, then
perhaps it is better to monitor, collect information and then act to scare vulnerable people sufficiently so that they do not get 
involved with any anticonstitutional group. But when should the government act? There is a trade-off here between
frustrating a number of attempts to radicalise vulnerable people, and bringing down a whole anticonstitutional group (with the attached risks
of failure and of exposing more susceptible individuals to malign influence for a longer period of time).
The decisions available to the radicalisers are similar; the asymmetry of the game arises from the fact that different decisions by both
players lead to very different collections of possible futures.

The Chain Event Graph (CEG) was introduced in 2008~\cite{PaulandJim} for the modelling of probabilistic problems whose underlying trees exhibit a high 
degree of asymmetry. It provides a platform from which to deduce dependence relationships between variables directly from the graph's
topology. CEGs have principally been used for learning/model selection (see for example~\cite{SilanderCEG,Lorna2}), but
also in two areas of interest to us in this paper -- causal analysis (see for example~\cite{CausalAI,RobandJim}), and also decision 
analysis~\cite{wupes} where the semantics of the CEG can be extended to provide
algorithms which allow users to discover minimal sets of variables needed to fully specify an SEUM decision rule.
In 2015 it was realised that CEGs include Acyclic Probabilistic Finite Automata (APFAs)
as a special case~\cite{EdwardsAPFA}.

In this paper we demonstrate how it is often possible to use causal CEGs to deduce (from appropriate qualitative assumptions) a simpler 
representation of a two person game. 
To retain plausibility we assume only the qualitative structure of the problem (as expressed by the topology of a CEG) is common knowledge, 
and that the players are SEUM given the information available to them when they make a move.
In section~2 we introduce the semantics of the decision CEG and 
discuss the principle of parsimony. To illustrate how the CEG can be used for the representation and analysis of games, and also how it can be used 
to simplify these games, section~3 contains a description of a 2~player game modelling a simplified version of the radicalisation scenario
described above.
Section~4 contains a discussion of ideas prompted by the work in earlier sections.

We have focussed here on two person adversarial games. However, similar techniques can be used both for non-adversarial games and also
for multi-player games. We have also assumed here that we are supporting one of the two players, but note that because of the common knowledge
assumption we have made, the qualitative results of the analysis are equally valid to this player's opponent or indeed some independent external observer.

\section{Decision Chain Event Graphs}	

\subsection{Conditional independence, Chain Event Graphs and causal hypotheses}	

Bayesian Networks (BNs) and Influence Diagrams express the conditional independence/Markov structure of a model through the presence/absence of edges 
between vertices of the graph. We say that a variable $X$ is independent of a variable $Y$ given $Z$ (written $X \amalg Y\ |\ Z$) if once we know
the value taken by $Z$, then $Y$ gives us no further information for forecasting $X$.
The structure of an ID can be used to produce fast algorithms for finding optimal decision strategies~\cite{Shachter86}.

One advantage that CEGs have over BNs and IDs for asymmetric problems is that they can be used to represent context-specific conditional independence
properties such as $X \amalg Y\ |\ (Z = z)$, which hold only for a subset of values of the conditioning variable.

The CEG is a function of a probability (or event) tree, having the same structure as a game tree, but with all non-leaf vertices being {\it chance} 
and all edges representing outcomes of these chance nodes, rather than actions of a player, We introduce two partitions of the vertices of the tree:
\begin{itemize}
\setlength{\itemsep}{-5pt}
\item[$\bullet$] Vertices in the same {\it stage} have sets of outgoing edges representing the same collections of possible outcomes, and have the
	same probabilities of these outcomes.
\item[$\bullet$] Vertices in the same {\it position} have sets of outgoing subpaths representing the same collections of possible complete futures, 
and have the same probabilities of these futures.
\end{itemize}

These equivalence classes encode (context-specific) conditional independences as follows: Given arrival at one of the vertices in a particular
stage, the next development is independent of precisely which vertex has been arrived at. Given arrival at one of the vertices in a particular
position, the complete future is independent of precisely which vertex has been arrived at. 

Our CEG is then produced from the tree by combining (or coalescing) vertices which are in the same position. Vertices in the same stage are generally 
given the same colour, and equivalent edges emanating from vertices in the same stage are generally also given the same colour. The stages and positions
between them encode the full conditional independence/Markov structure of our model. More detailed definitions are given in~\cite{PaulandJim}.

In~\cite{Pearl2000}, Pearl discusses the assumptions under which BNs can be considered causal (a more decision-theoretic
approach to graphical modelling is considered in~\cite{Jimbook}). We have shown that under similar
assumptions CEGs can also be considered as causal~\cite{CausalAI}. Heuristically this means that the model specified by a CEG continues to 
be valid when particular variables are manipulated. Such a hypothesis is a particularly natural one to entertain in decision problems, where a
decision maker (DM) by choosing a specific action at some point can be thought of as manipulating a specific variable. The hypothesis is also a
natural one to entertain in a game whose underlying structure is common knowledge and where
each player is able to manipulate their own decisions to a particular value,
but nature or the player's opponents will determine the value of other variables.

Those vertices in a CEG which we allow to be manipulated can be construed as decision nodes and the CEG as
a function of a decision tree \cite{JandPboth}. The
remaining vertices in the CEG are then chance or utility nodes. In this
mode the CEG is an elegant answer to the problems highlighted in~\cite{SHM,QZP,CandM,BandShen}, 
where different actions can result
in different choices in the future. As such it provides an alternative to valuation networks~\cite{Shenoy96},
decision circuits~\cite{BhatandS2}, sequential decision diagrams~\cite{CovandOli} and sequential IDs~\cite{JNandShen}. A full discussion of why 
IDs (including those supplemented by trees or similar) are unsatisfactory for the representation and analysis of asymmetric problems 
can be found in~\cite{BandShen}. A brief comparison of decision CEGs with IDs, and with valuation networks, decision circuits, sequential decision 
diagrams and sequential IDs can be found in~\cite{wupes}. A more detailed comparison will soon be available in an extended version of this paper.

In~\cite{CausalAI} we were concerned primarily
with the effects of a manipulation and whether these effects could be gauged
from probabilities in the idle system. Considering the CEG as a function of
a decision tree in contrast, we assume that the owner of the CEG
has a utility function over the possible outcomes of the problem, and can
then use techniques analogous to those used for decision trees to find an
optimal decision rule for the decision maker.
An introduction to the use of CEGs for decision analysis can be found in~\cite{wupes}. 

\subsection{An example of how CEGs can be used to represent decision problems}	

Figures~1 and~2 illustrate how a decision tree is converted into a CEG.
Figure~1 shows a variation of the Oil drilling
example from \cite{Raiffabook}. A more detailed version of this example appears in~\cite{wupes}.

\begin{figure}[ht]
\begin{center}
\includegraphics[height=3in]{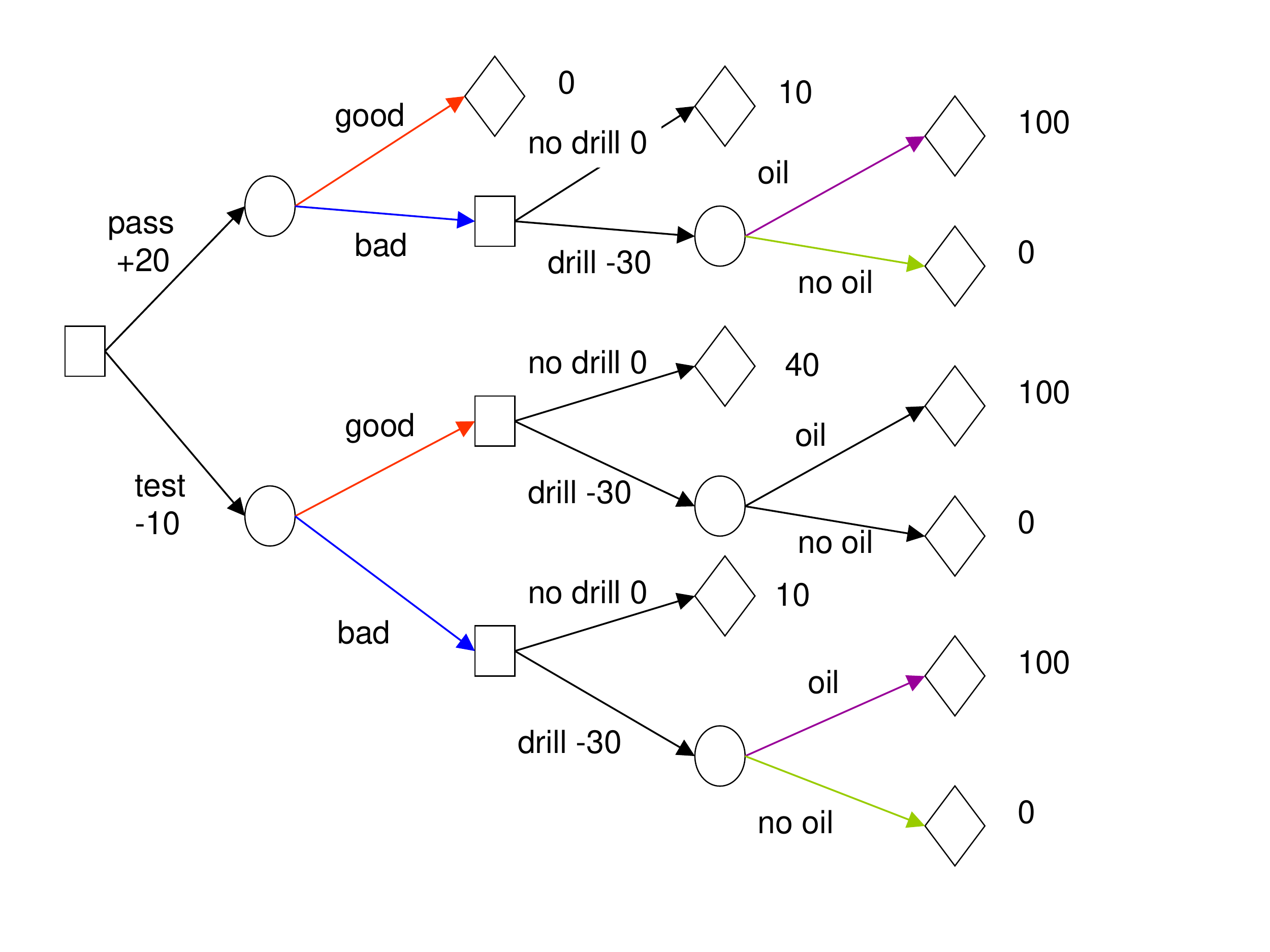}
\end{center}
\vspace{-10mm}
\caption{Coloured decision tree for Oil drilling example}
\label{fig:1}
\end{figure}

\begin{quotation}
\noindent\textit{We have an option on testing some ground for oil. We can
either  take up this option, at a cost of 10, or pass it on to another
prospector for a fee of 20.  Whoever does the testing, the outcomes are 
\emph{good} or \emph{bad}, with  probabilities independent of who does the
testing. If we have  passed on the testing and the test result is \emph{good}
then we lose the  option for drilling and get nothing. If it is \emph{bad}
then the other prospector  will not drill and the option for drilling
reverts to  us. If we do the test and the result is \emph{good}, then we can
either  drill, for a cost of 30, or sell the drilling rights for a fee of 
40. If the result is \emph{bad}, then regardless of who does the test,  we
can either drill ourselves, again for a cost of 30, or sell the  drilling
option for a fee of 10. If we drill and find oil we gain  100.}
\end{quotation}

\smallskip 
Decision nodes are indicated by squares, chance nodes by circles
and utility nodes by diamonds. Utilities here are decomposed and appear both on
the leaf utility nodes and on the edges emanating from decision nodes. Edges emanating from chance nodes have been assigned the same colour if the
nodes are in the same stage, and if the edges represent the same outcome and carry the same probability. So for example the two 
\textit{good}/\textit{bad} chance nodes are in the same stage. Also the first and third \textit{drill}/\textit{no drill} decision nodes are in the
same position, since they root subtrees which are identical in both topology and colouring. 

In purely-probabilistic CEGs
we usually combine all leaf nodes into a single sink node. With decision CEGs it is preferable to retain a number of leaf utility nodes, each representing
a different final reward. The CEG corresponding to Figure~1 is given in Figure~2. The CEG is Extensive Form -- that is the variables appear in the 
order that they appear to the DM. If there is oil in the ground it will have been there before we test or drill, but we don't know 
this until we drill. 

\begin{figure}[ht]
\begin{center}
\includegraphics[height=3in]{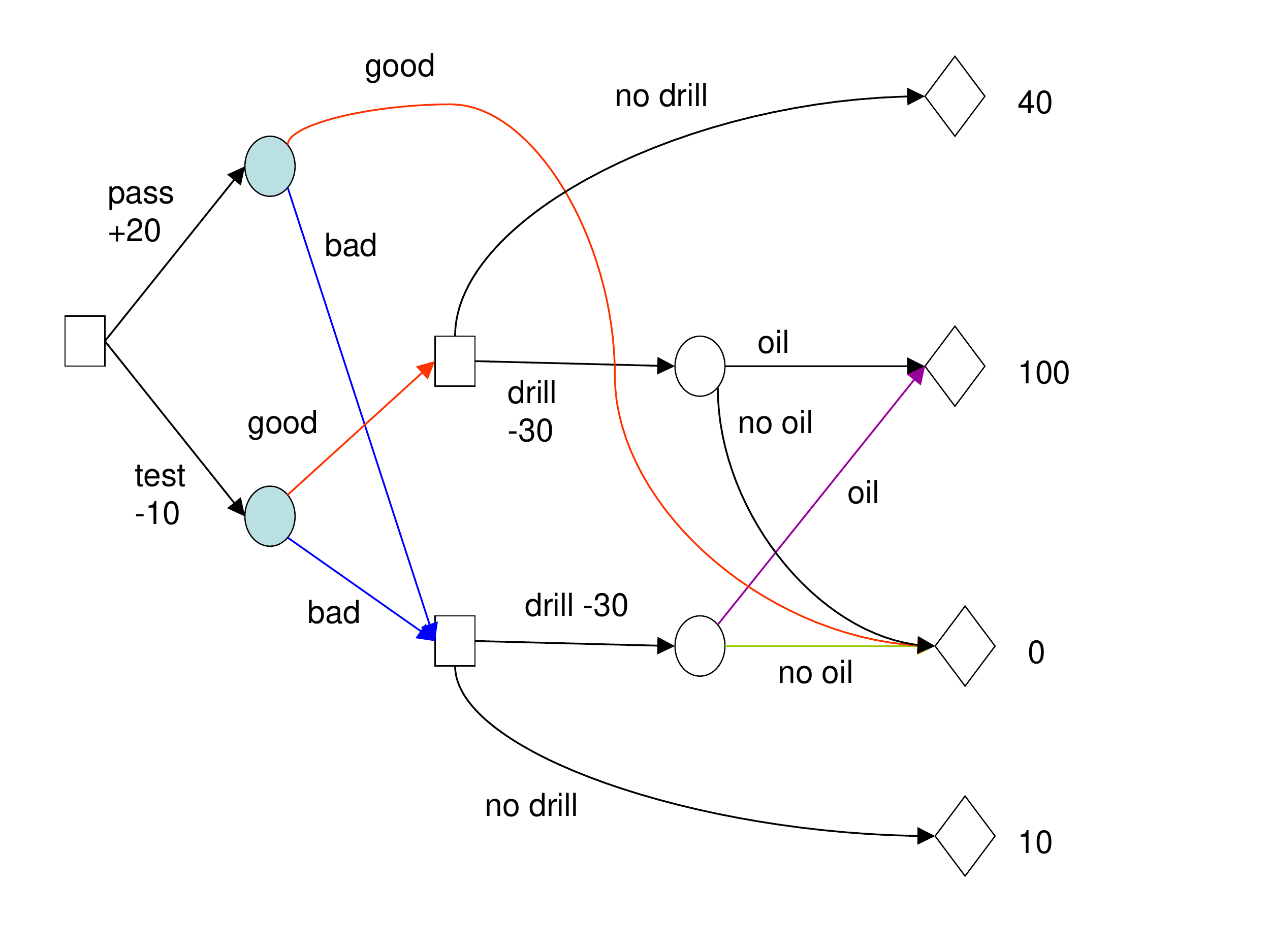}
\end{center}
\vspace{-10mm}
\caption{Decision CEG for Oil drilling example}
\label{fig:2}
\end{figure}

As with purely-probabilistic CEGs, the colouring and coalescence in the decision CEG allow it to express the complete conditional independence/Markov 
structure of a problem through its topology~\cite{wupes}. If vertices are in the same stage then we know that the possible immediate future developments
from these vertices are the same, and (if the vertices are chance nodes) that the probability of any specific immediate future development is the
same. Vertices in the tree which are in the same position are combined in the decision CEG because the sets of possible complete future developments from
these vertices are the same and have the same probability distribution.

In Figure 2 we have coloured the chance nodes which are in the same stage, and the edges emanating from them (indicating that they have the
same probabilities). We have retained the colouring of the edges leaving the 2nd {\em oil} chance node to illustrate how these edges relate to those
in Figure~1.

\subsection{Parsimony}	

In the example above we have a single DM. When we move into games with more than one player,
the CEG represents the problem to each of the players and its
topology can be considered as common knowledge (CK). The underlying
structure is causal to each player, but each decision node can only be
manipulated by a single (specified) player. In~\cite{Plausible} we
considered IDs which obeyed the same
assumptions as those described immediately above. Such IDs were resurrected
and modified by \cite{KolandM} and called MAIDs (multi-agent influence
diagrams). In \cite{Jim1989} we noted that such IDs could be seen from the
point of view of an informed observer to whom all nodes could be considered
as chance nodes -- a scenario which for example would be valid if the
observer's BN were causal and common knowledge. 

The paper~\cite{Plausible} describes how to produce a parsimonious representation of a game.
When a player needs to make a decision at $D$, some of the information they have obtained
beforehand may be superfluous for the purposes of making this decision. We then write 
\begin{align}
&U\amalg Q^{S}(D)\ |\ (D,Q^{R}(D))
\end{align}
where $U$ is the utility for that player, and $Q(D)=\{Q^{S}(D),Q^{R}(D)\}$ ($%
S=$ superfluous, $R=$ required) is a partition of the information obtained
by the player before making the decision (equivalent to the vertices in the
ID with edges directed into $D$). In this case the player need only consider the
configuration of the variables $Q^{R}(D)$ when choosing a decision at $D$ to maximise $U$. In \cite{KolandM}
the authors use similar ideas which they call \textit{strategic relevance} and \textit{s-reachability}. Conditional independence statements 
involving decision variables and utilities are discussed in some detail
in~\cite{wupes}. If the symmetry property of such statements is abandoned and a statement such as $X \amalg Y\ |\ Z$ is read only as {\it $X$ is
independent of $Y$ given $Z$}, then statements involving decision variables and utilities are unambiguous providing 
they do not take the form $D \amalg \dots\ |\ \dots$, where $D$ is a decision variable; and if the statement involves a utility $U$ then it
must be of the form $U \amalg \dots\ |\ \dots$.

\section{CEGs for Games, and an example of a 2 player game}	

In section~3.1 we introduce an example of a 2~player adversarial game, which we will use to illustrate how CEGs can be used to represent and analyse
such games. The ID of this example (Figure~3) appeared originally in~\cite{Plausible}
where it was used to demonstrate how the parsimony assumption can simplify
the analysis required in a 2~player game. The problem is presented here within the context of a test of strength between a government department and 
a radicalising website provider. In section~3.3 we demonstrate that the idea of parsimony can be
used to simplify CEG-based analysis; we also show how CEG methods
accomodate problem asymmetry in their topology in a way that IDs do not. Moreover
we see how the process of simplification can be directly linked to the
asymmetries exhibited in the associated game tree. 

\subsection{Example: description and ID}

In our example the players are the provider ($A$) of an internet site aimed at radicalising vulnerable people, and a government department (or police
force) tasked with combatting radicalisation ($B$). The site provider has contacts with a radical group (RG); the government department is aware of 
this group, but has no wish to tackle them directly at the present time. Our example is a simplification of the real games being played, and combines
aspects of both {\it prevention} and {\it pursuit} to illustrate how such games may develop. The example is concerned also with a single vulnerable
person (VP). As already noted, we could adapt the methods described below to games with more than two players, but here, for simplicity, 
the behaviour of VP and RG are considered to be governed by chance.

The ID contains chance and decision nodes as described below. We assume here that we are supporting one player. In the description which follows, there
is a {\it utility pair} associated with the chance variables $X_3$ and $X_4$, one of whose entries records our supported player's utilities, and the
other entry our supported player's estimated values for their opponent's utilities. For illustrative purposes, let our supported player be $A$ (!),
so the 2nd entry in our utility pairs is $A$'s estimate of $B$'s utility. In Figure~3, and in the CEGs in Figures 4 to~7, we have outlined $A$'s
decision nodes in red, and $B$'s decision nodes in blue, for ease of reading.

\begin{figure}[ht]
\begin{center}
\includegraphics[height=3in]{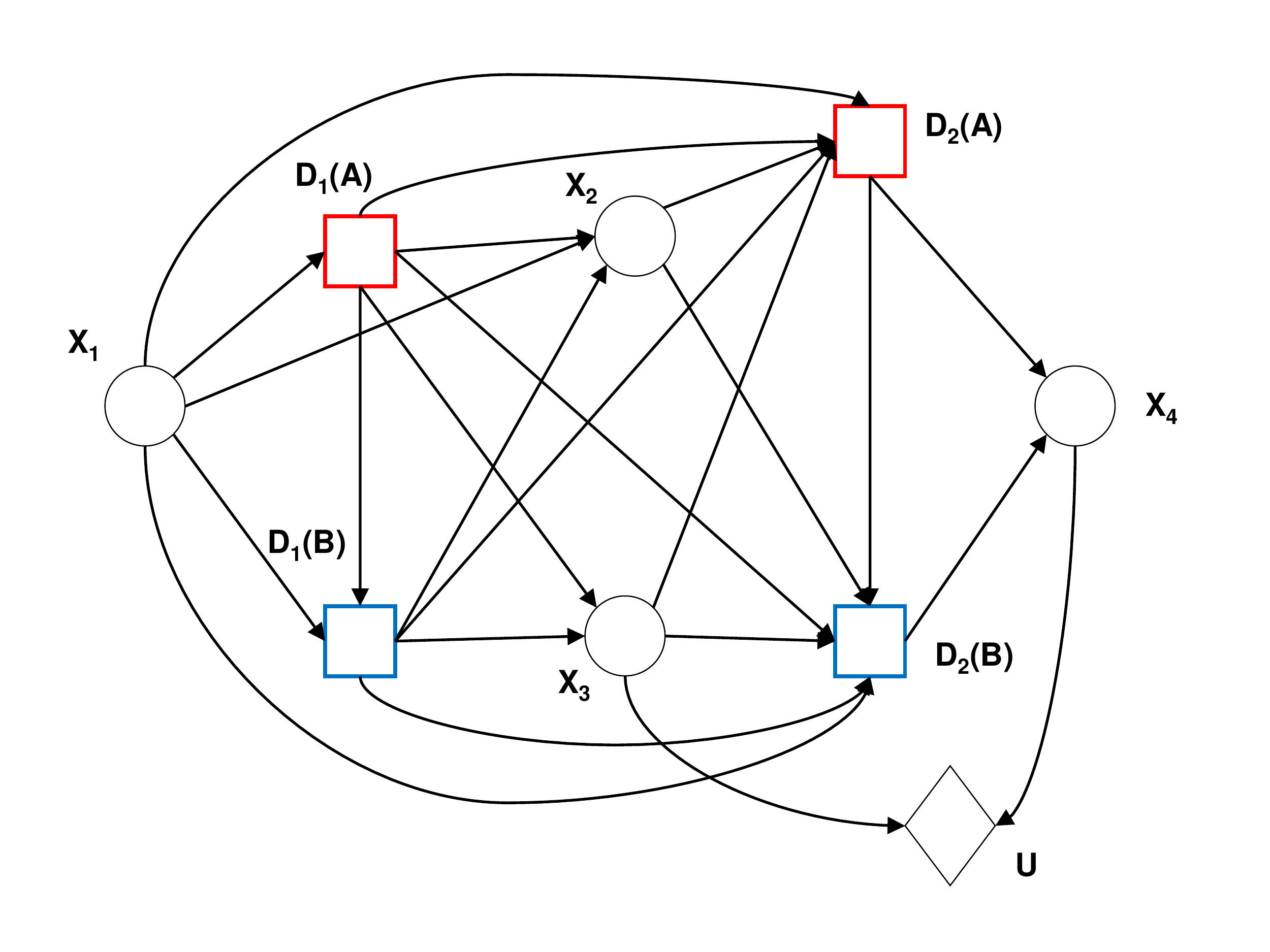}
\end{center}
\vspace{-10mm}
\caption{Initial ID}
\label{fig:3}
\end{figure}

\begin{itemize}
\item[$X_1$:] VP visits the website and either posts to the site, or contacts RG via a link on the site. Both $B$ \& $A$ observe VP's action.
\item[$D_1(A)$:] $A$ decides either to contact VP, or to contact RG. $B$ observes $A$'s action.
\item[$D_1(B)$:] $B$ decides either to contact VP via the site in the guise of an RG-sympathiser, or to shut the site down. $A$ observes $B$'s action.
\item[$X_2:$] $B$'s action at $D_1(B)$ may warn VP that he is being observed. The probability of this depends on whether VP posted to the site or 
	contacted RG, on whether $A$ contacted VP at $D_1(A)$ or not, and on $B$'s action at $D_1(B)$.\hfill\break
	VP's online behaviour indicating whether he is aware of being observed is itself observed by both $A$ \& $B$.
\item[$X_3$:] $A$'s action at $D_1(A)$ and $B$'s action at $D_1(B)$ either persuades RG to cut contact with $A$ (with utility pair $(U_A, U_B) = (-10, +10)$), 
	or to increase their cooperation (with utility pair $(U_A, U_B) = (+10, 0)$). The respective probabilities depend on which combination of $A$'s \& 
	$B$'s actions occurred. \hfill\break
	Behaviour of RG observed by both $A$ \& $B$.
\item[$D_2(A)$:] $A$ posts to own site either pretending to be VP, or in the guise of a sympathiser. This post provides false information about 
	themselves and their relationships with RG \& VP. VP knows that he did not post the message and that the aspect of the information concerning 
	his relationship with $A$ is false. The post is seen by $B$.
\item[$D_2(B)$:] $B$ decides either to arrest VP or not.
\item[$X_4:$] VP either tells $B$ that the information is false, or does not. The probability of VP doing this depends on $A$'s action at $D_2(A)$ and 
	$B$'s action at $D_2(B)$.
\end{itemize}

CEGs were designed for use with asymmetric problems. ID-representations often obscure such asymmetries and so even if a problem
is expressed as an ID it might still incorporate significant numbers of hidden asymmetries. A decision CEG can depict explicitly any
number of asymmetries, but for illustrative convenience we concentrate here on just one such possibility:

\begin{itemize}
\item[$\quad$] If at $X_3$, RG cut contact, then $A$ believes that $B$ believes that the information is false ($(U_A, U_B) = (0, 0)$) irrespective 
	of what VP tells $B$. 
	If at $X_3$, RG increased cooperation, then $A$ believes that $B$ believes that the information is true ($(U_A, U_B) = (+20, +10)$) if VP does 
	not tell $B$ that 
	it is false, and believes that it is false ($(U_A, U_B) = (0, 0)$) if VP tells $B$ it is false. 
\end{itemize}

\noindent{Note that $A$'s estimate of $B$'s utility for ($X_3$: RG increased cooperation, $X_4$: VP does not tell $B$ that information is false) is 
positive because $A$ believes that $B$ believes 
the information. $U_A$ is large for this scenario because $A$ believes that they have successfully planted false information on $B$.
If RG cuts contact at $X_3$ then the decisions made at $D_2(A), D_2(B)$ have no influence on $(U_A, U_B)$, and neither does the outcome of $X_4$.}

\begin{table}[h]
\caption{$(U_A, U_B)$ as a function of $X_3$ and $X_4$}
\begin{center}
\begin{tabular}{clcc}
&  &$X_4$: VP tells $B$ that false & VP does not ... \\ \hline 
$X_3$: &RG cuts contact & (-10, +10) & (-10, +10) \\ 
&RG increases ... & (+10, 0) & (+30, +10)%
\end{tabular}%
\end{center}
\end{table}

We describe in sections~3.2 and~3.3 how this information can be incorporated directly into the topology of the CEG.

\subsection{CEG and conditional independence structure}	

CEGs used for game analysis have the same relationship to game trees as Decision CEGs have to decision trees. 
In our CEG in Figure 4 there are no utilities on edges; they have been restricted to the terminal utility nodes (we have elsewhere described this 
as a Type 2 decision CEG~\cite{wupes}). For each utility node, each player has an associated utility pair (as described in section~3.1). From the
player's perspective, one value in the pair corresponds to their own utility value for this outcome, whereas the other corresponds to the player's
estimate of their opponent's utility value for this outcome. We maintain that the CEG in Figure~4 is causal, and that its
topology is common knowledge.

\begin{figure}[ht]
\begin{center}
\includegraphics[height=3.5in]{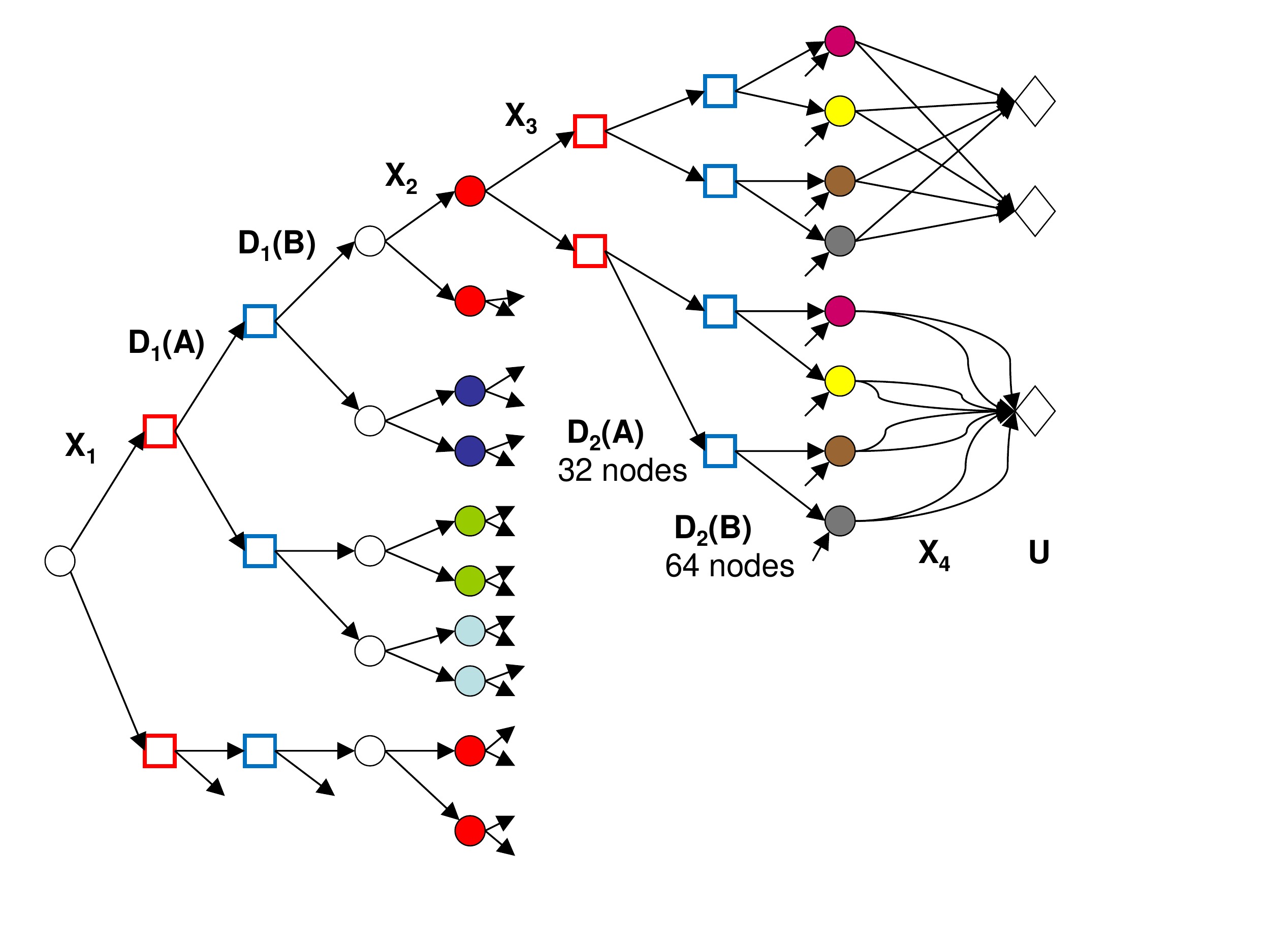}
\end{center}
\vspace{-10mm}
\caption{Naive CEG of model depicted in Figure 3}
\label{fig:4}
\end{figure}

We have given vertices in the same stage the same colour (as in Figure~2), but have not coloured emanating edges to avoid
cluttering the diagrams. We can read the colouring as, for example {\em the probability that $X_3 = 1$ given the histories up to each red $X_3$ vertex is the 
same, but different from that given the histories up to any of the differently coloured $X_3$ vertices}.

For illustrative convenience the CEG has been drawn out in more
detail than is strictly necessary. The diagram need really only show certain key aspects of the model -- the colouring of the $X_{3}$
vertices; the grouping and colouring of the $X_{4}$ vertices; and the asymmetric grouping
of the utility nodes (reflecting the utilities given in Table~1). These four aspects correspond to the four
non-trivial conditional independence statements associated with the model. So the full CEG need never be drawn out -- it can
simply be stored as a collection of computer constraints. The players need the picture only as a reminder of the key local properties and asymmetric
aspects of the game.

Various conditional independence/Markov properties can be read off CEGs by considering individual positions, stages or {\it cuts} through these. Stages 
encode statements about the immediate future, whereas positions encode statements about the complete future.

If we consider a cut through the 16 $X_3$ vertices, we see that they are grouped into four stages. So for instance the 1st, 2nd, 9th \& 10th $X_3$ 
vertices are in the same stage -- the probability that $X_3 = 1$ (say) is the same for the four histories $X_1 = 1\ \hbox{or}\ 2, D_1(A) = 1, D_1(B) = 1,
X_2 = 1\ \hbox{or}\ 2$.
Similar results hold for the other groups of vertices, and together give us the property that
$$X_{3} \amalg (X_{1},X_{2})\ |\ (D_{1}(A),D_{1}(B)).$$
This property concerns $X_3$, but not $D_2(A), D_2(B), X_4$ or $U$. It can also of course be read from the ID in Figure~3.
Similarly we see that the $X_4$ vertices are also grouped into four stages, and a similar reading of the stage cut through these vertices gives us the 
property
$$X_4 \amalg (X_1, D_1(A), D_1(B), X_2, X_3)\ |\ (D_2(A), D_2(B)).$$

\noindent{There are 8 $X_4$ vertices (although 4 stages, and in the underlying tree 128 vertices) because the utility function depends on the value taken by 
$X_4$ but also on the value taken by $X_3$ (which has no direct influence on $X_4$).}

When considering {\it position} cuts we ignore the colouring and groupings into stages. The first $X_4$ vertex corresponds to the histories 
$X_1 = 1\ \hbox{or}\ 2,\break D_1(A) = 1\ \hbox{or}\ 2, D_1(B) = 1\ \hbox{or}\ 2, X_2 = 1\ \hbox{or}\ 2, X_3 = 1, D_2(A) = 1, D_2(B) =1$. So reading the 
position cut through the $X_4$ vertices gives us the property
\begin{align}
&(X_4, U) \amalg (X_1, D_1(A), D_1(B), X_2)\ |\ (X_3, D_2(A), D_2(B)).
\end{align}
Here the property concerns $X_4$, but also $U$, since positions encode statements about the complete future.
We finally consider a cut through the utility nodes, and see that
$$U \amalg (X_1, D_1(A), D_1(B), X_2, D_2(A), D_2(B))\ |\ (X_3, X_4).$$
Our experience suggests that with practice, users of CEGs quickly become adept at the reading of the graphs for their conditional independence structure.

\subsection{Simplifying the CEG}	

We now turn our attention to how parsimony allows us to simplify analysis.
For a multi-player adversarial game, simplification takes the form of an iterative process whose steps are of two types -- decision node coalescence,
and barren node deletion. The process is run from leaf nodes to root node.

\noindent{{\bf Decision node coalescence:} 
As already noted, for a player making a decision at $D$, if she can learn that 
$U\amalg Q^{S}(D)\ |\ (D,Q^{R}(D))$, then she need only consider the
configuration of the variables $Q^{R}(D)$ when choosing a decision at $D$ to maximise her utility. 
If $Q^S(D)$ is non-empty then in the CEG there are distinct decision nodes that are actually in the same position and 
therefore can be combined~\cite{wupes}.
Two or more decision nodes (in a Type~2 decision CEG) are in the same position if:}
\begin{itemize}
\setlength{\itemsep}{-5pt}
\item the subCEGs rooted in these nodes have the same topology,
\item equivalent edges in these subCEGs have the same labels and (where appropriate) probabilities,
\item equivalent branches in these subCEGs terminate in the same utility node.
\end{itemize}

\noindent{{\bf Barren node deletion:} As with IDs~\cite{Shachter86}, decision CEGs may have barren nodes which can be deleted. A barren node in a 
Type~2 decision CEG~\cite{wupes} is simply a 
vertex all of whose emanating edges terminate in the same node. If the vertex is a decision node then whatever decision the DM makes is irrelevant, and 
if it is a chance node then whatever outcome happens is of no consequence. The deletion step proceeds as follows~\cite{wupes}:} 

If $w$ has only one child node then
\begin{itemize}
\setlength{\itemsep}{-5pt}
\item label this node $w_{\succ}$,
\item for each node $w_{\prec}$ in the parent set of $w$:\hfill\break replace all edges $e(w_{\prec}, w)$ by a single edge $e(w_{\prec}, w_{\succ})$,
	and delete all edges $e(w, w_{\succ})$ and the node $w$.
\end{itemize}

We have noted that this simplification process works for adversarial games with two or more players. We expect that it will also work,  
possibly with some minor modification, for non-adversarial games as well. 

With these tools at our disposal we can start to simplify the CEG from Figure~4.

The 5th to 8th $X_4$ nodes in Figure~4 are barren, as all their emanating edges terminate in the same utility node. They can be deleted, and the
edges entering them from $D_2(B)$ vertices extended so that they now terminate at the third utility node.
But these $D_2(B)$ nodes are now also barren, as all their emanating edges now terminate in the same utility node. They can be deleted, and the
edges entering them from $D_2(A)$ vertices extended so that they now terminate at the third utility node.
These $D_2(A)$ nodes are now also barren, and once we have deleted them we get a reduced graph as in Figure 5.

\begin{figure}[ht]
\begin{center}
\includegraphics[height=3.5in]{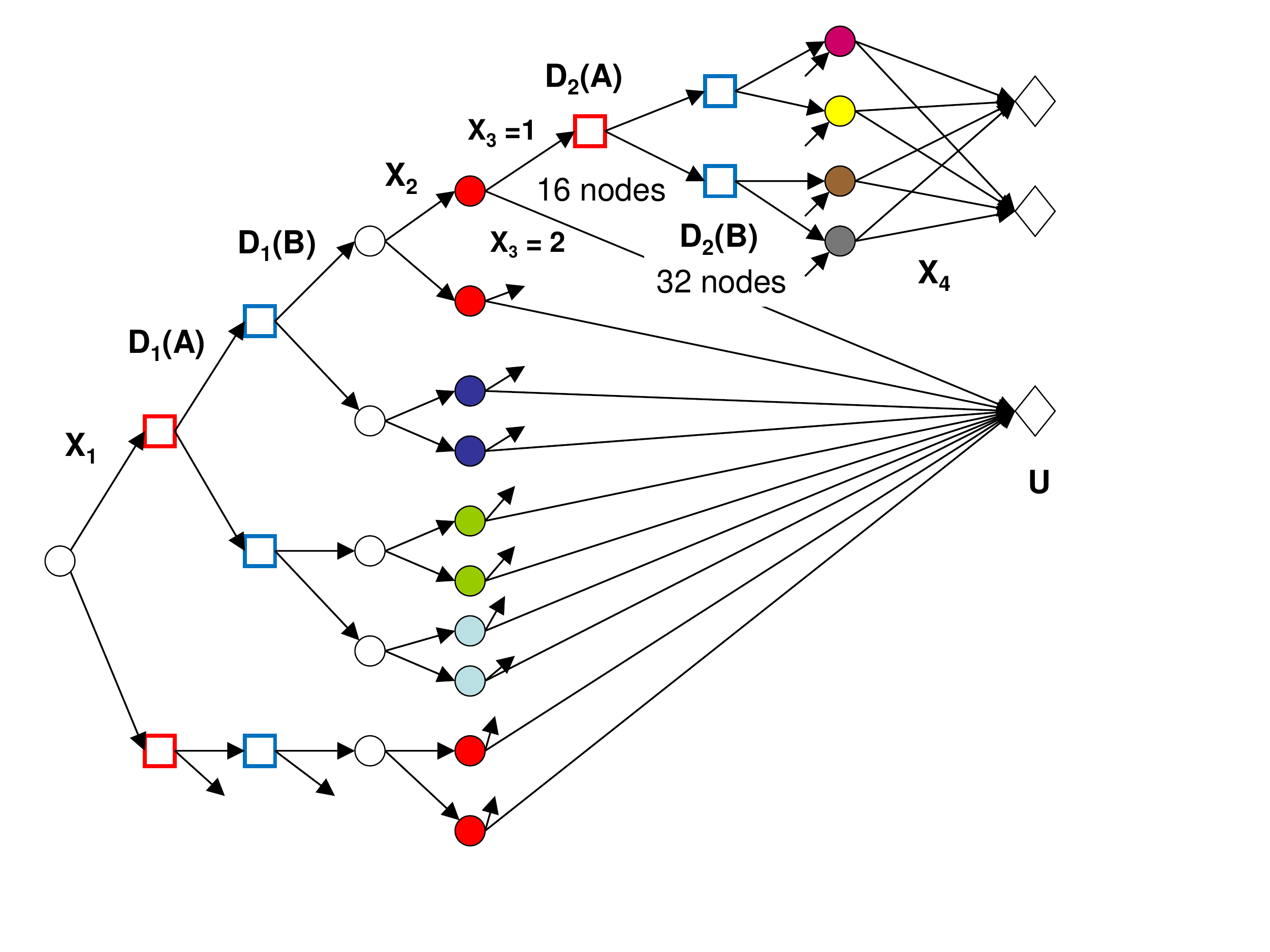}
\vspace{-10mm}
\caption{First reduced CEG} \label{fig:5}
\end{center}
\end{figure}

Now considering the remainder of the graph, we can deduce from expression~(2) that 
\begin{align}
&U_B \amalg (X_1, D_1(A), D_1(B), X_2)\ |\ (D_2(B), X_3 = 1, D_2(A)).
\end{align}
So $Q^S(D_2(B)) = \{X_1, D_1(A), D_1(B), X_2 \}$, and these variables can be considered as non-parents of $D_2(B)$ for the purposes of optimal decision
making. The 32 remaining $D_2(B)$ vertices are grouped into two positions (corresponding to the combinations $X_3 = 1, D_2(A) = 1\ \hbox{or}\ 2$), and we
can combine these $D_2(B)$ vertices into two, each with 16 incoming edges corresponding to the 16 possible configurations of $(X_1,
D_1(A), D_1(B), X_2)$. Each remaining $X_4$ vertex now has only 1 incoming edge.

A position cut through the $D_2(B)$ nodes now yields the statement
$$(D_2(B), X_4, U) \amalg (X_1, D_1(A), D_1(B), X_2)\ |\ (X_3 = 1, D_2(A)),$$
which if viewed from the perspective of $B$ is an invalid statement -- $B$ can choose arbitrarily between two actions at $D_2(B)$,
so the statement that $D_2(B)$ is conditionally independent of other variables is nonsensical (see comment in section~2.3).

\begin{figure}[ht]
\begin{center}
\includegraphics[height=3.5in]{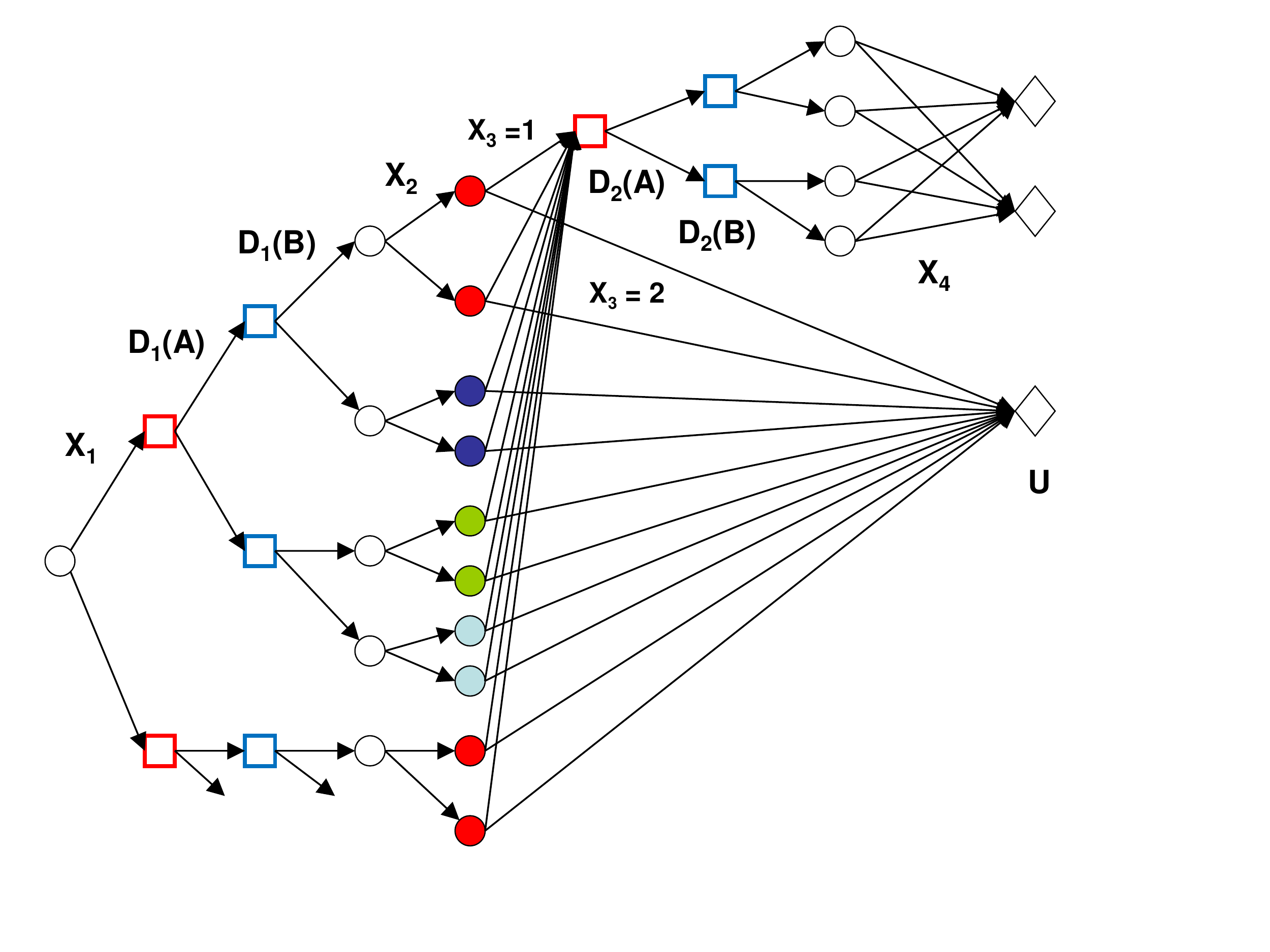}
\vspace{-10mm}
\caption{Second reduced CEG} \label{fig:6}
\end{center}
\end{figure}


However, this is a 2 player game, and at this point in the analysis we are considering the possible actions of $A$ at $D_2(A)$. To $A$ the
behaviour of $B$ can be considered as random (or at least $D_2(B)$ can be considered as a chance variable), and so to $A$ the statement
has meaning. We can therefore deduce that
$$U_A \amalg (X_1, D_1(A), D_1(B), X_2)\ |\ (D_2(A), X_3 = 1).$$
So $Q^S(D_2(A)) = \{X_1, D_1(A), D_1(B), X_2 \}$, and these variables can be considered as non-parents of $D_2(A)$ for the purposes of optimal decision
making. The 16 remaining $D_2(A)$ vertices are all in the same position (corresponding to $X_3 = 1$), and we
can combine these $D_2(A)$ vertices into one, with 16 incoming edges corresponding to the 16 possible configurations of $(X_1,
D_1(A), D_1(B), X_2)$. The two remaining $D_2(B)$ vertices now each have only 1 incoming edge. The resulting graph is given in Figure~6, where the
redundant colouring of the remaining $X_4$ vertices has been removed.


We cover the remainder of the simplification process more rapidly. The 1st, 2nd, 9th \& 10th $X_3$ vertices root identical subCEGs, so they are now 
in the same position and can 
be combined. The same is true of the other sets of coloured $X_3$ vertices.	
The 8 $X_2$ nodes are barren and can be deleted (we run the edges from the 4 $D_1(B)$ vertices straight into the appropriate $X_3$ vertices).
The 1st \& 3rd $D_1(B)$ vertices now root identical subCEGs so are in the same position. The same is true of the 2nd \& 4th $D_1(B)$ vertices, and 
the two $D_1(A)$ vertices.
Finally, the $X_1$ node is now barren so can be removed, giving us the {\it parsimonious} CEG in Figure~7. 

\begin{figure}[ht]
\begin{center}
\includegraphics[height=3.5in]{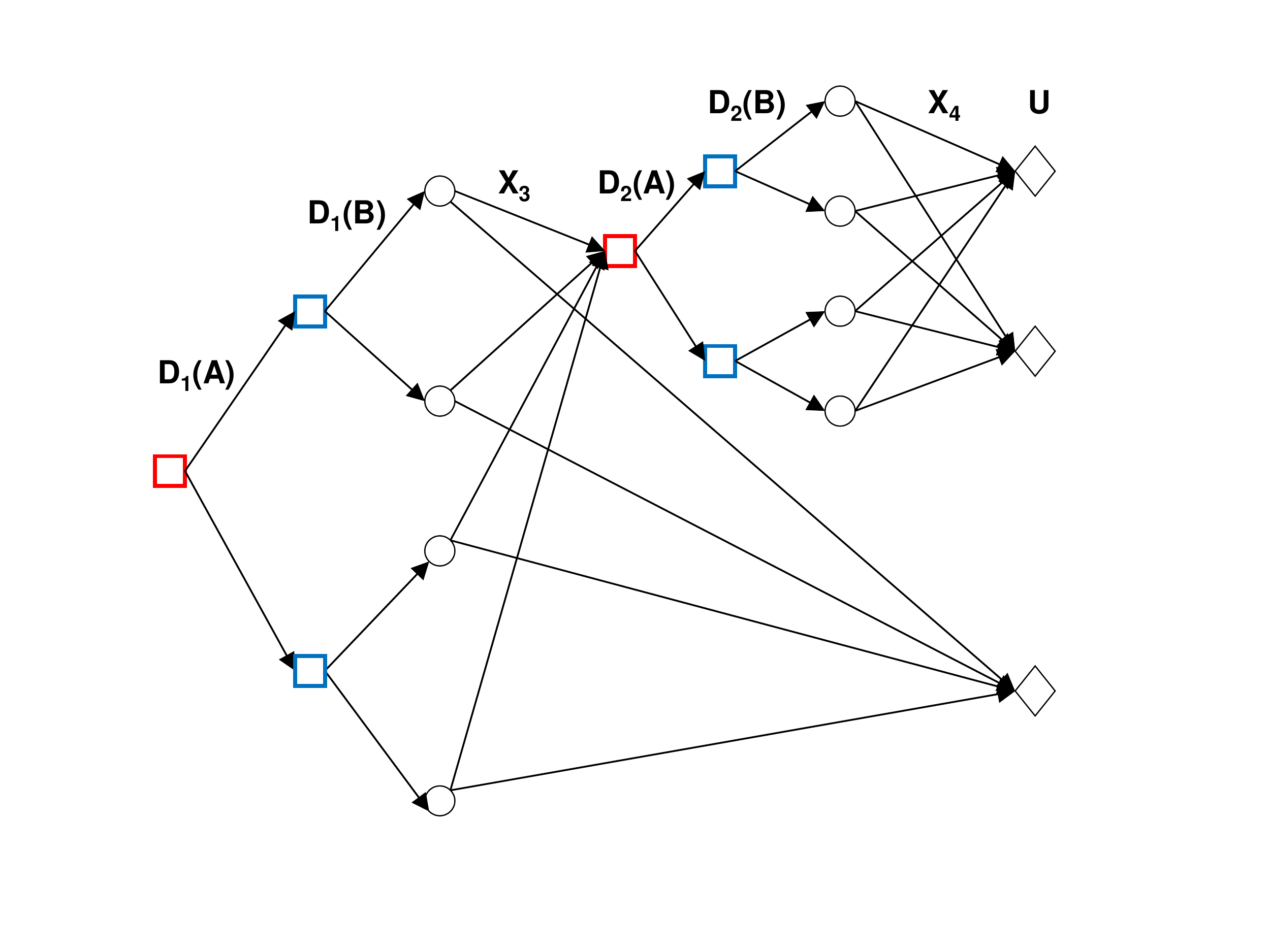}
\vspace{-10mm}
\caption{Parsimonious CEG} \label{fig:7}
\end{center}
\end{figure}

At this point it is worth reminding ourselves that we are supporting one player, here $A$, so even when we have considered $U_B$ (as in expression~(3)),
we have been looking at the game from $A$'s perspective. Now, both players start with the same initial CEG (Figure~4) since its topology is
considered to be common knowledge; and some aspects of the simplification process will occur regardless of which player's shoulder we are looking
over (such as the removal of the vertices associated with $X_1$ and $X_2$). But other aspects of the simplification might be different for $B$, since 
the process will depend on $B$'s own utilities and on $B$'s estimates of $A$'s utilities, rather than on $A$'s beliefs about 
$B$'s utilities and $A$'s own utilities. The particular shape of the parsimonious CEG in Figure~7 is a result of $A$'s utility pair being $(0, 0)$ if 
RG cuts contact, irrespective of what VP tells $B$. But $B$ may not have the same utility for both possible cases here, and may not believe
that $A$ has the same utility for both. So $B$ might produce a different parsimonious CEG to $A$, although still simpler than the original CEG.

We know that $X_1$ and $X_2$ are irrelevant
for optimal decision making purposes (for both $A$ and $B$). From Figure~7 we see that from $A$'s perspective, $X_3$ depends on both $D_1(A)$ and 
$D_1(B)$. If $X_3 = 2$ then $A$ believes that decisions made at $D_2(A)$, $D_2(B)$
are irrelevant. Only if $X_3 = 1$ are they of any consequence and $A$ believes that they do not need to consider any other prior action or event when 
making a decision at $D_2(A)$. In this case $B$ needs to consider $D_2(A)$ when making a decision at $D_2(B)$. $A$'s own utility and $A$'s estimated 
utility for $B$ depend only on the values taken by $X_3$ and~$X_4$. 

The parsimonious CEG in Figure~7 is not much more complex than the equivalent ID, which has the
vertices $D_1(A), D_1(B), X_3, D_2(A), D_2(B)$ \& $U$, and the edges 
$D_1(A) \rightarrow D_1(B), X_3;\ D_1(B) \rightarrow X_3;\ X_3 \rightarrow D_2(A), D_2(B), U;\ 
D_2(A) \rightarrow D_2(B), X_4;\ D_2(B) \rightarrow X_4$ and $X_4 \rightarrow U$~\cite{Plausible}. Koller and Milch, comparing MAIDs and game
trees in~\cite{KolandM}, note that {\it a MAID representation is not always compact. If a game tree is naturally asymmetric, a naive MAID representation
can be exponentially larger than the tree}.
The parsimonious ID does not of course encode any
of the asymmetry of the problem, a major drawback when evaluating optimal decision rules for the players.

\subsection{Solution}	

The most appropriate solution concept for a game expressed as Extensive Form with Chance moves is Bayes-Nash equilibrium. The equilibria can
be found by applying a decision analysis type {\it rollback} or backward induction algorithm to the game tree or CEG, in which each player plays a 
{\it best response} to the strategy of the other player(s).

In our game, our players are SEUM, conditioned on the information available to them each time they make a decision, and hence they are {\it sequentially
rational}. Consequently, our backward induction computes subgame perfect equilibria, and the end result of the process is a subgame perfect Nash
equilibrium~\cite{Banks}.

As noted by Banks et al in~\cite{Banks}, if our game incorporates any sort of asymmetry (such as that described in this paper), then attempts to
deduce Nash equilibria from non-tree based representations of the game (such as {\it pay-off tables}) will usually yield impossible equilibria.
This happens because representations which assume symmetry in the game will provide utility pairs (or in general, utility vectors) for combinations of
decisions which could not possibly happen.

Now this process will produce agreed equilibria for the game if each player's utilities are common knowledge. But the process is still valid if 
this is not so, and we are supporting one player. The equilibria will simply be those that our supported player believes exist, based on her own 
utilities and her estimates of the utilities of the other players.

Returning to our example, once the qualitative re-analysis of section~3.3 is complete, the optimal decision rule for $A$ and the decision rule which
$A$ believes to be optimal for $B$ (and which $A$ therefore thinks $B$ will follow) can be discovered  by treating the CEG from Figure~7 as if
there were a single decision maker, working upstream from the terminal nodes,
but noting that when we reach the set of $D_{2}(B)$ vertices, the optimal decisions will be those that maximise $A$'s estimates of $B$'s utility, 
and that when we reach the $D_{2}(A)$ vertex, the optimal decision will be that which
maximises $A$'s utility and so on. An algorithm for this process is given in Table~2.

In the algorithm we use $C$ and $D$ for the sets of chance and decision nodes, with $D(B)$ indicating a decision node belonging to $B$ (etc).
$U_A[w]$ and $(U_A, U_B)[w]$ (etc) indicate $A$'s utility at the position $w$
and the utility pair at $w$ (etc). The (conditional) probability of an edge $e(w,w')$ is denoted by $p[e(w,w')]$. The set of child nodes of a position~$w$ 
is denoted by $ch(w)$. Note that  $\sum_{e(w_i,w)}$ appears in the line ``If $w_i \in C$" because there may be more than one edge connecting two 
positions, if say two different decisions have the same consequence. 

Once the algorithm has run, the root node will have an associated utility pair $(U_A, U_B)[w_1]$, 
such that $U_A[w_1]$ will be $A$'s maximum expected utility given their assumptions, and $U_B[w_1]$ will be $B$'s expected
utility if they follow the strategy that $A$ believes they should. $A$'s optimal decision strategy and $B$'s strategy if $A$ is correct in their
beliefs about $B$ will be indicated by the subset of edges that have not been marked as being sub-optimal. 

\begin{table}[ht]
\hrule
\caption{Local propagation algorithm for finding a subgame perfect Nash equilibrium} 
\begin{center}
\begin{flushleft}%
\begin{itemize}
\setlength{\itemsep}{-5pt}
\item Find an ordering of the positions $w_1, w_2,\ldots, w_n$, such that 
\begin{enumerate}
	\item $w_1$ is the root-node, 
	\item if there are $t$ utility nodes, then $w_{n-t+1}, w_{n-t+2}, \dots , w_n$ are the utility nodes,
	\item if $w_j \in ch(w_i)$ then $j > i$.
\end{enumerate}
\item Initialize the utility pairs $(U_A, U_B)[w]$ of the utility leaf nodes to the values specified.
\item Iterate: for $i=n-t$ step minus 1 until $i=1$ do:
\begin{itemize}
\item If $w_i \in C$ then\hfill\break $(U_A, U_B)[w_i] := \sum_{w \in ch(w_i)}\big\{\sum_{e(w_i,w)}\big\{ p[e(w_i,w)]*(U_A, U_B)[w]\big\}\big\}$
\item If $w_i \in D(B)$ then $(U_A, U_B)[w_i] := (U_A, U_B)[w^*]$\hfill\break where $w^* = \arg\max_{w \in ch(w_i)}\big\{U_B[w]\big\}$ 
\item If $w_i \in D(A)$ then $(U_A, U_B)[w_i] := (U_A, U_B)[w^*]$\hfill\break where $w^* = \arg\max_{w \in ch(w_i)}\big\{U_A[w]\big\}$ 
\end{itemize}
\item Mark the sub-optimal edges.
\end{itemize}
\end{flushleft}%
\end{center}
\hrule
\end{table}

For our example, noting that $X_3 = 1$ corresponds to RG {\it increases contact}, Table~1 gives us $A$'s utility pairs for the 3 terminal utility nodes. 
They are $(U_A, U_B) = (+10, 0), (+30, +10)$ and $(-10, +10)$.

If at $D_2(A)$, $A$ has chosen $D_2(A) = 1$ (the upper edge emanating from the $D_2(A)$ vertex), then at $D_2(B)$, $A$ believes that
$B$ needs to choose a decision $D_2(B) = 1\ \hbox{or}\ 2$ based on whichever of
\begin{align*}
&p (X_4 = 1\ |\ X_3 = 1, D_2(A) = 1, D_2(B) = 1) \times 0 \\
+\ &p (X_4 = 2\ |\ X_3 = 1, D_2(A) = 1, D_2(B) = 1) \times +10 \\
\hbox{and}\hskip2cm &p (X_4 = 1\ |\ X_3 = 1, D_2(A) = 1, D_2(B) = 2) \times 0 \\
+\ &p (X_4 = 2\ |\ X_3 = 1, D_2(A) = 1, D_2(B) = 2) \times +10
\end{align*}
is greater. A similar expression exists for if $A$ has chosen $D_2(A) = 2$.

In deciding on an action at $D_2(A)$, $A$ assumes that $B$ will act rationally at $D_2(B)$. If we denote by $d^1_2(B)$ \& $d^2_2(B)$ the rational
decisions of $B$ (as perceived by $A$) at $D_2(B)$ given that $A$ chose $D_2(A) = 1$ or $D_2(A) = 2$, then $A$ needs to choose a decision 
$D_2(A) = 1\ \hbox{or}\ 2$ based on whichever of
\begin{align*}
&p (X_4 = 1\ |\ X_3 = 1, D_2(A) = 1, D_2(B) = d^1_2(B)) \times +10 \\
+\ &p (X_4 = 2\ |\ X_3 = 1, D_2(A) = 1, D_2(B) = d^1_2(B)) \times +30 \\
\hbox{and}\hskip2cm &p (X_4 = 1\ |\ X_3 = 1, D_2(A) = 1, D_2(B) = d^2_2(B)) \times +10 \\
+\ &p (X_4 = 2\ |\ X_3 = 1, D_2(A) = 1, D_2(B) = d^2_2(B)) \times +30
\end{align*}
is greater. Similar decision rules can be generated for $B$ at each $D_1(B)$ vertex and for $A$ at $D_1(A)$.

From $A$'s perspective, $B$'s decisions at the $D_2(B)$ vertices reduce to choosing the action which will lead to the higher conditional probability 
that $X_4 = 2$; and there
are similar simple interpretations of decisions at other points in the graph, so even if the players have no software for processing the information 
stored in the graph, they can make their choices very quickly.

Starting with the CEG in Figure~4 and populating it with $B$'s own utilities and beliefs about $A$'s utilities, $B$ can produce their own
simplified CEG from which they can discover their own optimal decision rule in an exactly analogous manner.

As with the simplification process described in section~3.3, the algorithm given in Table~2 can be easily adapted for use with multi-player
adversarial games. Non-adversarial games may require something slightly more complex.

\section{Discussion}

The example in section~3 illustrates how Bayesian game theory can be used in constructing models of competitive
environments based on the impact the likely rationality of players might have on the \textit{structure} of the game. Once the class
of models consistent with this structure has been identified, we can use standard Bayesian techniques to estimate its
parameters. Bayesian game theory can thus be used to enhance and complement a Bayesian analysis, making it more plausible from a perspective of mutual
rationality. 

Structural reasoning, such as common knowledge assumptions and the idea of
parsimony, gives ways to deduce simplified forms of the players' decision rules. Distributions for our supported player's opponent can then be elicited,
based, for example, on their previous acts and what our supported player believes the distribtuions of the outcome
variables to be. This then gives a standard decision CEG to solve, but one that recognises the more plausible structural common knowledge and is fashioned
to be consistent with this. Alternatively, in the special case where the players really
do believe that they can assess their opponent's utility function accurately and
both players agree on the probability assignment,
then we can simply proceed in a standard game-theoretic way: we add
this extra information to the common knowledge base, and seamlessly apply
this to compute a solution of the game.
Note that in this case our prior structural analysis has helped because it
has prevented us engaging in elicitation activities which subsequently prove to be superfluous

Our example is obviously a simplification of the real game played between governments and radicalisers, but is sufficiently detailed both to show
how the asymmetry powers the analysis, and describe in essence how such games function. 
One thing that the many similar real games have in common is that at a population level there are people who if caught
\& dealt with early, will never become involved in anything anticonstitutional again, and those for whom it really is a game, and for whom
such interventional methods will not work. The constitutional organisation needs to decide whether to concentrate on the former group of people
(on the grounds that they will get greater reward for doing so) or the latter group (who perpetuate the game). Their utility will be a function of the risks 
inherent in either strategy.

As noted in section~3.1, our example could be thought of as a four player game. If we interpret it as such, we would need our supported player~$A$
to consider the utilities also of VP and RG. We would also have decision nodes associated with both VP and RG. We 
have confined ourselves to two players here simply for illustrative convenience, and to demonstrate the power and simplicity of the method. It would
be straightforward to extend the methodology to three, four or more players. Little modification is needed -- the graph is simplified following the same rules,
and in the rollback required for maximum expected utility calculations, the utilities to be maximised will (still) be those of whichever player
has to make a decision at that point.

We have also confined ourselves here to an adversarial game (of the sort described in the introduction), 
but we believe there is scope for adapting the ideas and techniques to the modelling
of other forms of games such as those involving oligopolistic competition. If we were to interpret the game in our example as a four player game,
then there would probably be some degree of cooperation or collusion between some of the players, particularly between $A$ and~RG. It is unlikely
that this would have much effect on the iterative simplification process described in section~3.3, but it would require some modification to
the solution algorithm of section~3.4. At $A$'s decision nodes, $A$ might wish to choose actions which maximised some function of their own 
utility and RG's utility, rather than just their own utility.

In our example both $A$'s and $B$'s utilities depended on the chance variables $X_3$ and$X_4$; and this was common knowledge. $B$'s
utilities could however depend on different variables to $A$'s, provided the dependence structure was common knowledge. In the ID-representation of the
game, this could be encoded by having separate utility nodes for $A$ and $B$. One way that this could be represented in a CEG without increasing
the topological complexity of the graph, would be to keep the single collection of terminal utility nodes, each representing a different utility 
pair; but read any conditional independence properties involving utilities on a player-by-player basis (as we did in the simplification process
in section~3.3). An alternative is to extend the CEG so that each $U_A$ utility node is connected by a single edge to a $U_B$ utility node. There
might then be $U_A$ utility nodes that are in the same stage but not the same position. This slight increase in complexity would be offset by
still being able to read the full conditional independence structure from the topology of the graph alone. Other possibilities exist.

The use of tree models in decision analysis waned during the ascendancy of ID-based representations and solution methods. Developments such as CEGs
and the types of graphical models described in~\cite{Jaeger04,JNandShen,BhatandS2}  etc make this once-again a particularly
powerful modelling tool for decision problems, and as we have shown here, for Bayesian games.

\bigskip
\noindent{{\bf Acknowledgements:} This research is being supported by the EPSRC -- project EP/M018687/1 {\it Modelling Decision
and Preference problems using Chain Event Graphs}. We would also like to express our thanks
to Robert Cowell for his input into the early development of the theory of decision CEGs, and to earlier versions of the example from section~2.}

\end{document}